# X-ray follow up observations of new IGRs

**Jerome Rodriguez**[*][†]

*Laboratoire AIM, CEA/IRFU - CNRS/INSU - Université Paris Diderot, CEA DSM/IRFU/SAp,
Centre de Saclay, F-91191 Gif-sur-Yvette, France*
*E-mail:* `jrodriguez@cea.fr`

**Arash Bodaghee & John Tomsick**

*Space Sciences Laboratory, 7 Gauss Way, University of California, Berkeley, CA 94720-7450,
USA*
*E-mail:* `bodaghee@ssl.berkeley.edu, jtomsick@ssl.berkeley.edu`

Since the launch of INTEGRAL in 2002, about 300 new sources have been discovered. Understanding the nature of these objects is of prime importance for many aspects of astrophysics, such as the evolution of stars, population of sources (Galactic and extra-Galactic), and ultimately the physics powering them. However, their nature cannot be established from the soft $\gamma$-ray observations. The first step towards unveiling the nature of those sources is to refine their X-ray position, in order to finally find counterparts at other wavelengths. X-ray spectra are also of prime importance to obtain clues on the nature of the objects. Since the discovery of the first IGR in 2003, our group has been active in several aspects of these studies. Here, we present the main results we have obtained through 7 years of multi-instrumental (Chandra, XMM, Swift, RXTE) campaigns.

---

[*]Speaker.
[†]A footnote may follow.

## 1. Introduction

The most recent version of the IBIS catalog contains more than 700 hard X-ray sources (Bird et al. 2010). While a certain number were known as (hard) X-ray emitters prior to the launch of *INTEGRAL*, about half of them have been detected for the first time above 20 keV with IBIS/ISGRI (Lebrun et al. 2003). In the remainder of this paper we will refer to these sources as 'IGRs'[1]. Bodaghee et al. (2007) collected known parameters (e.g., the absorption column density, $N_H$, the pulse period for Galactic sources with X-ray pulsations, the redshift for AGN, etc.) of all sources detected by *INTEGRAL* during the first four years of activity. With this they could study the parameters spaces occupied by different families of sources and therefore could deduce important aspect concerning the physics of high energy sources.

Today, however, many of these IGRs are still unidentified. Understanding the nature of these objects is of prime importance for many aspects of astrophysics, such as the evolution of stars, population of sources (Galactic and extra-Galactic), and ultimately the physics powering them. It is, however, not possible to identify those sources from soft γ-ray observations. One first needs to refine the position of the objects using soft X-ray telescopes such as *Chandra*, *XMM-Newton*, or *Swift*. Only after this steps has been performed can one browse the optical/infrared catalogues or perform observations with optical/IR telescope to search for counterparts at longer wavelengths, and try to unveil their nature. In parallel a lot can be learnt from the study of the joint soft-hard X-ray spectra of those sources, as well as from temporal analysis.

Since 2003, our team has been quite active both in several aspects of these studies, starting with X-ray follow-up observations with all available facilities: *Chandra* (Tomsick et al. 2006, 2008, 2009), *XMM-Newton* (Rodriguez et al. 2003, 2006; Bodaghee et al. 2006), *Swift* (Rodriguez et al. 2008, 2009b, 2010) or even *RXTE* for the search for specific temporal features (Rodriguez et al. 2009a).

In this paper we illustrate these different aspects by showing some of the main results we have obtained through 7 years of multi-instrumental campaigns. In section 2 we present the identification that has been made possible by a refinement of X-ray astrometry, and search for optical/IR counterparts. In Section 3 we show how X-ray spectral and temporal analysis only can be useful in determining the nature of an X-ray object, while in Section 4 we present a recent multi-instrumental/multi-wavelength analysis that permitted us to discover a distant, powerful and rather young pulsar wind nebula.

## 2. X-ray astrometry and search for counterparts

Refining the X-ray position can be a first and easy step towards understanding the nature of a given source. Sometimes, indeed, the sky is quite cooperative, and the refined X-ray position is coincident with the nucleus of a galaxy. In those obvious cases, the galactic nucleus and the X-ray emission permit to classify the source as AGN. Fig. 1 shows an example of one of these "easy" cases. Here the *Swift* refined position allowed us to identify NGC 4939 as the counterpart to IGR J13042-1020, and thus conclude that the source is an AGN (Rodriguez et al. 2010). Examination of the X-ray spectra further enabled us to designate the source as a probable Type 2 Seyfert AGN, while browsing online catalogues we could find a distance z$\sim$ 0.01 for this object. Thanks to Swift we have identified 22 AGNs, 1 Bl Lac, 5 HMXBs, 2 CVs, 5 X-Ray binaries, and 1 Rs CVn (Rodriguez et al. 2008, 2009, 2010). Obvious cases are not the majority, even with the Chandra astrometry. In most cases the sources and their counterparts are point-like and only photometric and spectrometric studies of the counterpart can allow the type of the object to be identified. Following these approaches, first based on the *Chandra* extremely precise astrometry, we could identify 17 HMXBs, 2 SNRs, 8 CVs, and 5 AGNs (Tomsick et al. 2006, 2008, 2009).

---

[1] An up-to-date on-line catalog of all IGRs can be found at http://irfu.cea.fr/Sap/IGR-Sources/ note the new address for the site

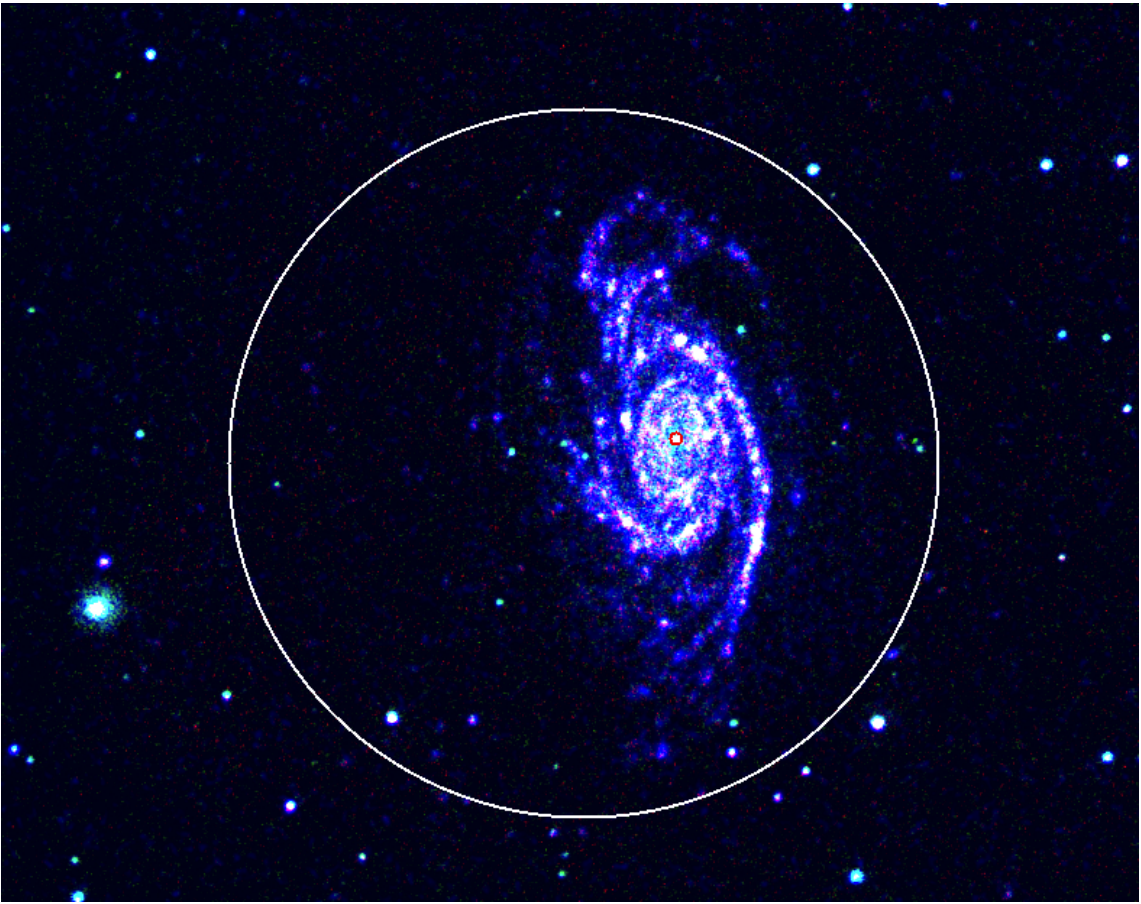

**Figure 1:** *Swift*/UVOT composite image of the field of IGR J13042-1020. The white circle is the IBIS error box, while the small red circle is the *Swift*/XRT error box. July 2010 INTEGRAL Picture of the Month adapted from Rodriguez et al. (2010).

## 3. Spectral and temporal X-ray studies

X-ray spectral and temporal analysis are two extremely powerful ways to determine the nature of a source (besides studying the physics powering them of course). Fig. 2 shows examples of one source (namely IGR J19294+1816)for which our spectro-temporal studies indicated the presence of an X-ray pulsars in an HMXB. Here the *RXTE* data permitted us to find coherent pulsations in the periodogram of the source (Rodriguez et al. 2009a). These pulsations are indicative of polar accretion onto a pulsar.The pulse period is thought to be the spin period of the pulsar. In other cases, as IGR J16320−4751, the spectral analysis of simultaneous XMM-INTEGRAL spectra showed the presence of different spectral components, such as an iron emission line, and local absorption at low energy of a Comptonised spectrum. We also discovered the presence of a soft excess in this source (Rodriguez et al. 2006). Time and phase-resolved spectral analysis of these data allowed us to conclude that IGR J16320−4751 is an HMXB.

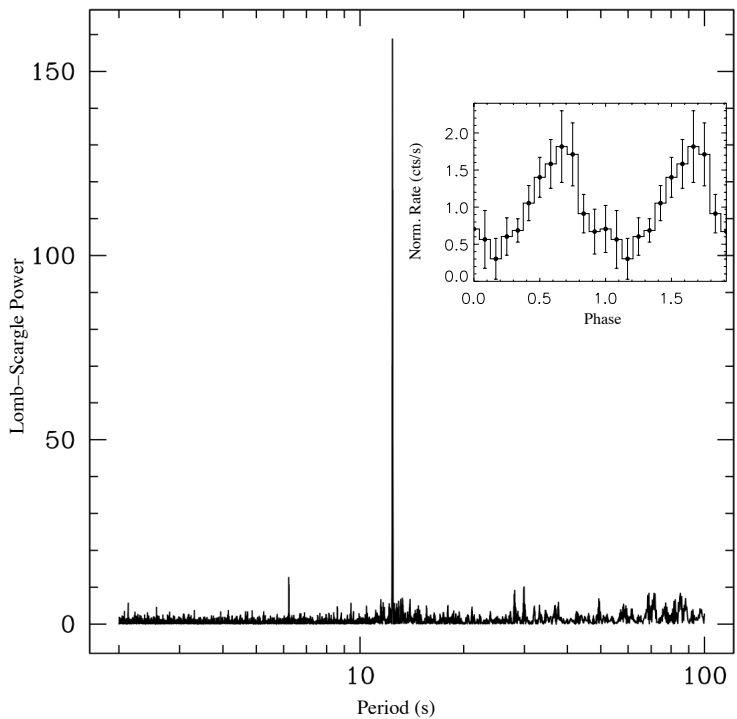

**Figure 2:** Periodogram of the *RXTE* Êobservation of IGR J19294+1816 showing a clear peak at 11.3s. The insert shows the light curve of the source folded at this period. Figure from Rodriguez et al. (2009a).

## 4. Multi-instrumental campaigns

Finally all these approaches can be put together to access more precisely the physics powering the sources. This is nicely exemplified by IGR J14003-6326 where our Chandra observation allowed us to suggest a possible PWN due to the extension of the source in the Chandra image (Tomsick et al. 2009, Fig. 3). Our RXTE observations have permitted us to find pulsations at 31.2 ms (Fig. reffig:14003), which was confirmed at radio wavelengths (Renaud et al. 2010). This pulse period makes the source one of the fastest rotators in the Galaxy. Through the analysis of the whole multi instrumental campaign, we concluded that IGR J14003-6326 is a powerful, distant and rather young object (Renaud et al. 2010).

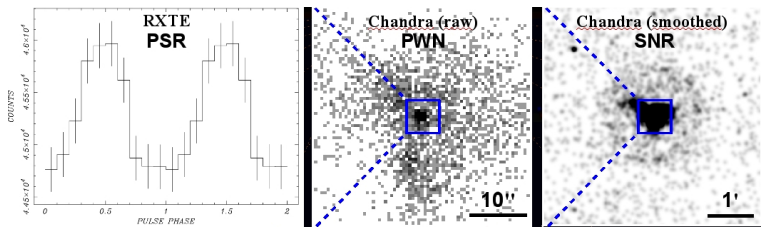

**Figure 3:** IGR J14003−6326 *Chandra* images and *RXTE* light curve folded at the 31.2 ms pulsation. Images adapted from Tomsick et al. (2009) and Renaud et al. (2010).

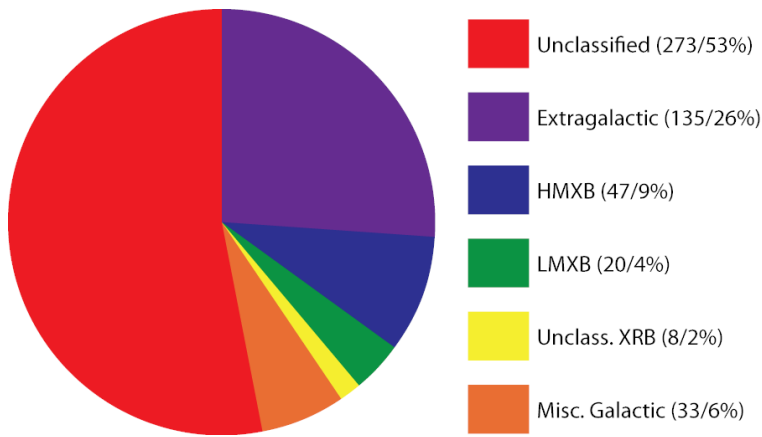

**Figure 4:** Distribution of the different type of sources amongst the IGRs.

## 5. IGR Source population

Fig. 4 shows the repartition of the current (as of November 2010) populations of IGR. It is obvious that a large fraction of IGRs is still unidentified, therefore follow ups need to be continued. Among identified sources AGN represent the majority, and this ratio should grow as the INTEGRAL observation cover more and more high Galactic latitudes.